\documentclass[11pt]{article}
\usepackage{moriond,epsfig}

\def\be{\begin{equation}}
\def\ee{\end{equation}}
\def\bea{\begin{eqnarray}}
\def\eea{\end{eqnarray}}

\newcommand{\met}       {\mbox{$\not\!\!E_T$}}
\newcommand{\rar}       {\rightarrow}
\newcommand{\ppbar}     {\mbox{$p\bar{p}$}}
\newcommand{\ttbar}     {\mbox{$t\bar{t}$}}
\newcommand{\rargap}    {\mbox{ $\rightarrow$ }}
\def\vtb{|V_{tb}|}
\begin{document}
\vspace*{4cm}
\title{Single top quark production at the Tevatron}

\author{ R. Schwienhorst\footnote{On behalf of the D0 and CDF collaborations.}  }

\address{Michigan State University, 3234BPS, East Lansing, MI 48824, USA}

\maketitle\abstracts{
The Tevatron experiments D0 and CDF have found evidence for single top quark
production, based on datasets between 0.9~fb$^{-1}$ and 2.2~fb$^{-1}$. Several
different multivariate techniques are used to extract the single top quark signal
out of the large backgrounds. The cross section measurements are also used to
provide the first direct measurement of the CKM matrix element $\vtb$.
}

\section{Introduction}
\begin{figure}
\begin{minipage}{0.5\textwidth}
  \centering
  \psfig{figure=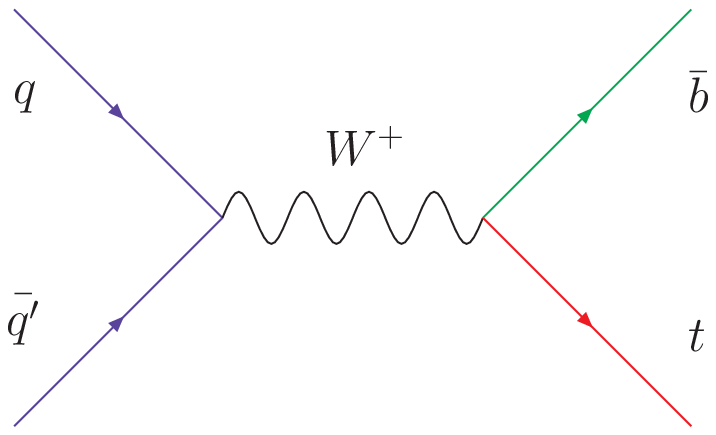,height=1.2in}
\end{minipage}
\begin{minipage}{0.5\textwidth}
  \centering
  \psfig{figure=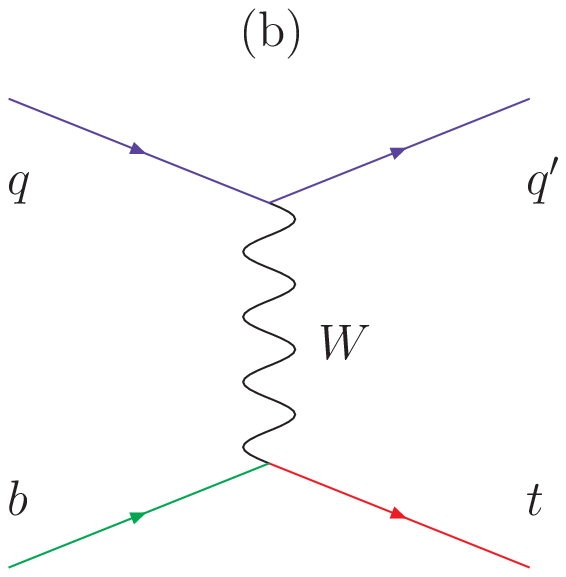,height=1.2in}
\end{minipage}
\caption{Feynman diagrams for s-channel
(left) and t-channel (right) single top quark production at the Tevatron.
\label{fig:feynmantbtqb}}
\end{figure}

Evidence for single top quark production at the Tevatron and a first direct measurement
of the CKM matrix element $\vtb$ was first reported by the
D0 collaboration~\cite{run2-d0-prl-evidence}. In contrast to top quark pair production
through the strong interaction, which was observed in 
1995~\cite{top-obs-1995-cdf,top-obs-1995-d0}, single top quarks are produced via the
weak interaction. 
The Feynman diagrams for standard model (SM) s-channel ($tb$) and t-channel ($tqb$) single top 
quark production are
shown in Fig.~\ref{fig:feynmantbtqb}. There is third production mode, associated production of
a top quark and a $W$ boson, but its cross section is so small that it will not be considered
further.
The SM cross section for the s-channel process ${\ppbar}{\rar}t\bar{b} + X, \bar{t}b +X$ 
is $0.88 \pm 0.14$~pb at NLO for 
$m_{\rm top} = 175$~GeV~\cite{singletop-xsec-sullivan,Cao:2004ap}. At the same
order and mass, the cross section
for the t-channel process ${\ppbar}{\rar}tq\bar{b} + X,\bar{t}\bar{q}b + X$ is
$1.98 \pm 0.30$~pb~\cite{singletop-xsec-sullivan,Cao:2005pq}.

Measuring the single top quark production cross section provides a direct measurement
of the CKM matrix element $\vtb$. 
The single top quark final state also allows for studies of the top quark polarization,
and it is sensitive to many models of new physics, for example flavor changing neutral
currents via the gluon~\cite{Abazov:2007ev} or heavy new bosons $W'$ that only couple
to quarks~\cite{Abazov:2008vj}. The s-channel process is also an important background to
Higgs searches in the associated production mode, and the advanced analysis techniques 
used in the single top searches will be applicable to Higgs searches as well.

The D0 collaboration has updated two of their analysis methods using a dataset of 
0.9~fb$^{-1}$. The updated results, including a combination of different methods 
are presented below.
The CDF collaboration has analyzed a dataset of 2.2~fb$^{-1}$ and significantly
improved the sensitivity to single top quark production. These new results
are presented below.

\section{D0 results}

\subsection{Event selection}
The D0 analysis selects electron+jets and muon+jets events in 0.9~fb$^{-1}$ of data 
with the following requirements:
One high-$p_T$ lepton (electron ($pT>15~GeV$) or muon ($p_T>18~GeV$)), missing transverse
energy $\met>15GeV$, and
between two and four jets with jet $p_T>15~GeV$ and jet~1 $p_T>25GeV$, at least one is tagged
with a neural-network based b-tagging algorithm. Additional cuts remove
fake-lepton background events. Events are collected by lepton+jets trigger
requirements.

The number of events observed in data and expected from the background model and 
SM signal is shown in Table~\ref{tab:d0-event-yields}. The largest sources of systematic
uncertainty are the background normalization, jet energy scale, as well as b-tag and trigger
modelling.
\begin{table}[th]
\caption[eventyields]{Numbers of events expected by D0 in 0.9~fb$^{-1}$ for
electron and muon, 1 $b$-tag and 2 $b$-tag channels combined.
\label{tab:d0-event-yields}}
\begin{center}
\begin{tabular}{| l | r@{$\pm$}l | r@{$\pm$}l | r@{$\pm$}l |}
\hline
                 &  \multicolumn{2}{|c|}{2 jets}
                 &  \multicolumn{2}{|c|}{3 jets}
                 &  \multicolumn{2}{|c|}{4 jets} \\
\hline
s-channel        &   16  &    3  &    7  &   2  &    2  &   1  \\
t-channel        &   20  &    4  &   12  &   3  &    4  &   1  \\
\hline
$\ttbar$         &   59  &   14  &  134  &  32  &  155  &  36  \\
$W$+jets         &  531  &  129  &  248  &  64  &   70  &  20  \\
Multijets        &   96  &   19  &   77  &  15  &   29  &   6  \\
\hline
Total background &  686  &  131  &  460  &  75  &  253  &  42  \\
Data             &  \multicolumn{2}{|c|}{697}
                 &  \multicolumn{2}{|c|}{455}
                 &  \multicolumn{2}{|c|}{246}      \\
\hline
\end{tabular}
\end{center}
\end{table}

Table~\ref{tab:d0-event-yields} shows that after selection cuts, the expected SM single top
signal is small compared to the background sum, and in fact the signal is significantly smaller
than the background uncertainty. Thus, more advanced techniques are required to extract the 
signal.

\subsection{Multivariate techniques}
The D0 analysis employs three different multivariate techniques to extract the single
top quark signal out of the large backgrounds. The boosted decision tree (BDT) analysis 
has not changed since the publication of evidence for single top quark 
production~\cite{Abazov:2007ev}. Here we focus on the Bayesian neutral network analysis 
and the matrix element analysis, both of which have been re-optimized.

\begin{figure}
\begin{minipage}{0.5\textwidth}
  \centering
  \psfig{figure=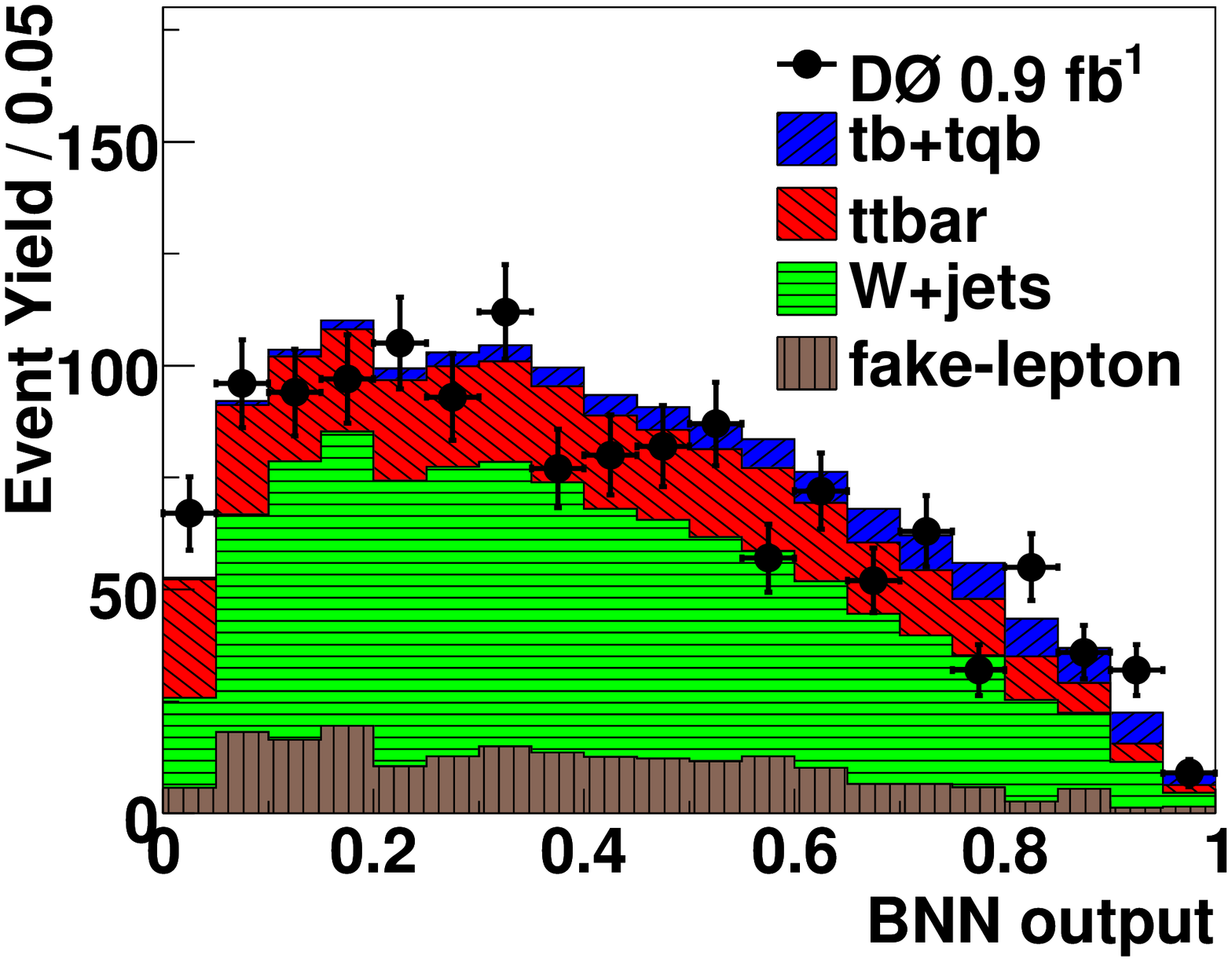,height=1.8in}
\end{minipage}
\begin{minipage}{0.5\textwidth}
  \centering
  \psfig{figure=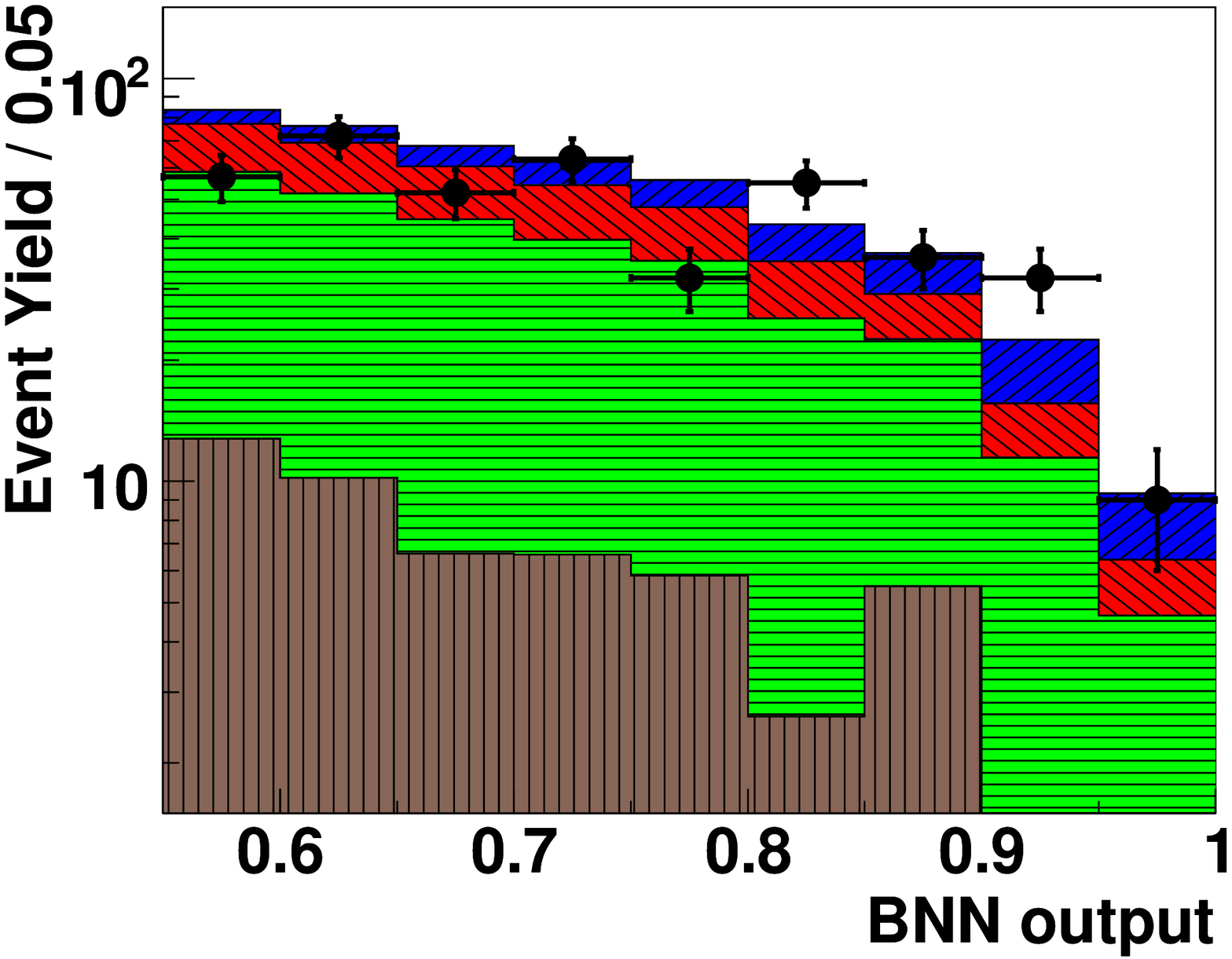,height=1.8in}
\end{minipage}
\caption{Comparison between data and background sum for the Bayesian neural network output. 
Shown is the full distribution (left), and the high-discriminant region (right). The signal
has been normalized to the SM expectation. 
\label{fig:d0-bnnoutput}}
\end{figure}

In a conventional neural network, the network parameters and weights are determined in
an optimization (training) procedure. Rather than optimizing for these network parameters
once and then fixing them, the optimal network configuration can be obtained as an
average over many different values for the network parameters. In this Bayesian
procedure, an integration over all of the possible network parameter space
is performed. The network architecture is fixed, and the weight of each set of parameters
is obtained through a Bayesian integration. The final network discriminant is then
the weighted average over all the individual networks. 
Fig.~\ref{fig:d0-bnnoutput} shows the output of the BNN for the D0 data.

The Matrix element analysis starts from the Feynman diagrams for the single top quark processes
and uses transfer functions to relate the parton level quark-level information to the reconstructed 
jet and other information. Matrix elements for the single top quark signal as well as the $W$+jets
backgrounds are included. For 3-jet events, a top pair matrix element is also included. For each 
event, an integration over the phase space is performed, employing the
transfer functions to compute the probability for this particular event to arise from a specific
matrix element. A likelihood function is then formed as the ratio of the signal and signal plus
background probabilities.

\subsection{D0 summary}
\begin{figure}
  \centering
  \psfig{figure=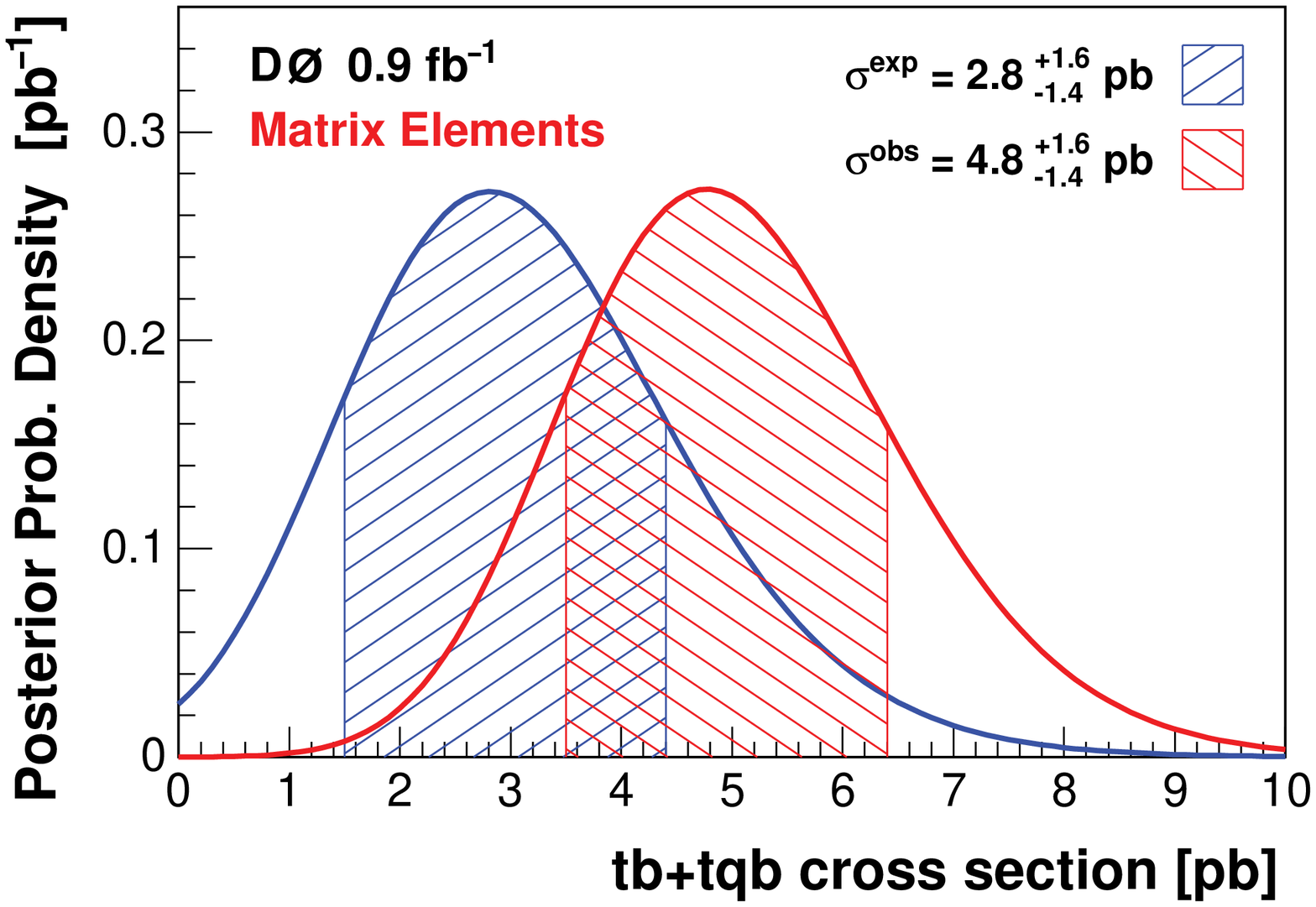,height=1.5in}
  \psfig{figure=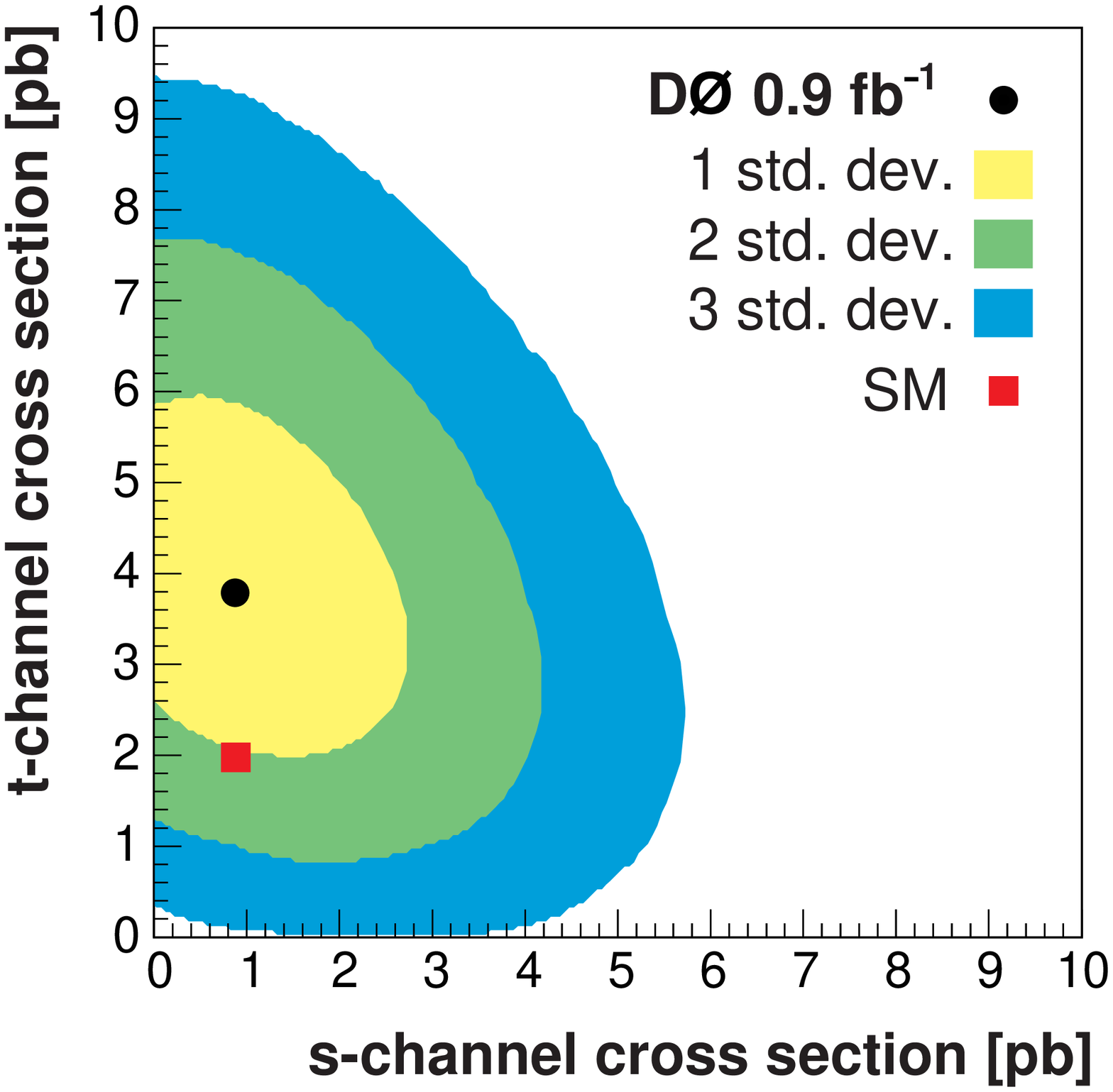,height=1.8in}
\caption{Posterior probability density for the matrix element analysis as a function of the sum 
of s-channel and t-channel cross sections (left), and for the BDT analysis as a function 
of both the s-channel and t-channel cross sections (right). 
\label{fig:d0-2dpost}}
\end{figure}

The cross section is measured as the peak of the Bayesian posterior probability density,
shown in Fig.~\ref{fig:d0-2dpost} for the ME analysis.
The three different methods measure the following cross sections for the sum of
s-~and t-channel:
$$
\begin{array}{llll}
 \sigma^{\rm obs}\left({\ppbar}{\rargap}tb+X,~tqb+X\right)
 & = & 4.9 ^{+1.4}_{-1.4}~{\rm pb} & {\rm (DT)}  \\
 & = & 4.4 ^{+1.6}_{-1.4}~{\rm pb} & {\rm (BNN)} \\
 & = & 4.8 ^{+1.6}_{-1.4}~{\rm pb} & {\rm (ME)}.
\end{array}
$$
The measured cross sections are consistent with each other and above the SM expectation.

The decision tree analysis has also measured the s- and t-channel cross sections
separately,
$$
\begin{array}{lll}
 \sigma^{\rm obs}\left({\ppbar}{\rargap}tb+X\right)
 & = & 1.0\pm 0.9~{\rm pb} \\
 \sigma^{\rm obs}\left({\ppbar}{\rargap}tqb+X\right)
 & = & 4.2^{+1.8}_{-1.4}~{\rm pb},
\end{array}
$$
where the standard model cross section is used for the single
top process not being measured.

Removing the constraint of the standard model ratio allows to
form the posterior probability density as a function of both the $tb$ and $tqb$
cross sections. This model-independent posterior is shown in
Fig.~\ref{fig:d0-2dpost}~(right)for the DT analysis, using the $tb$+$tqb$
discriminant. The most probable value corresponds to cross sections of
$\sigma(tb)=0.9$~pb and $\sigma(tqb)=3.8$~pb. Also shown are the one,
two, and three standard deviation contours. While this result favors a
higher value for the $t$-channel contribution than the SM expectation,
the difference is not statistically significant. Several models of new
physics that are also consistent with this result are shown in
Ref.~\cite{beyond-SM-tait}.
These updated results have recently been published~\cite{Abazov:2008kt}.

\section{CDF results}

\subsection{Event selection}
The CDF analysis selects electron+jets and muon+jets events in 2.2~fb$^{-1}$ of data
with the following requirements:
One high-$p_T$ lepton ($p_T>20~GeV$), $\met>25GeV$, and
two or three jets with jet $p_T>20~GeV$, at least one of which
is tagged by a displaced vertex tagging algorithm. 
Additional cuts remove fake-lepton background events. Events are collected by single-lepton
trigger requirements. The matrix element analysis uses
additional triggers in the muon channel to increase the acceptance.

The number of events observed in data and expected from the background model and 
SM signal is shown in Table~\ref{tab:cdf-event-yields}. The largest sources of systematic
uncertainty are the background normalization, jet energy scale, and b-tag modelling.
\begin{table}[th]
\caption[eventyields]{Numbers of events expected by CDF in 2.2~fb$^{-1}$ for
electron and muon, 1 $b$-tag and 2 $b$-tag channels combined.}
\label{tab:cdf-event-yields}
\begin{center}
\begin{tabular}{| l | r@{$\pm$}l | r@{$\pm$}l |}
\hline
                 &  \multicolumn{2}{|c|}{2 jets}
                 &  \multicolumn{2}{|c|}{3 jets} \\
\hline
s-channel        &   41  &    6  &   14  &   2    \\
t-channel        &   62  &    9  &   18  &   3    \\
\hline
$\ttbar$         &  146  &   21  &  339  &  48    \\
$W$+bottom       &  462  &  139  &  141  &  43    \\
$W$+charm        &  395  &  122  &  109  &  34    \\
$W$+light        &  340  &   56  &  102  &  17    \\
$Z$+jets         &   27  &    4  &   11  &  2    \\
diboson          &   63  &    6  &   22  &  2    \\
Multijets        &   60  &   24  &   21  &  9    \\
\hline
Total background & 1492  &  269  &  755  &  91    \\
Data             &  \multicolumn{2}{|c|}{1535}
                 &  \multicolumn{2}{|c|}{752} \\
\hline 
\end{tabular}
\end{center}
\end{table}
Again, it is clear that a advanced analysis techniques are required to extract the signal.

\subsection{CDF Likelihood Function}
\begin{figure}
  \centering
  \psfig{figure=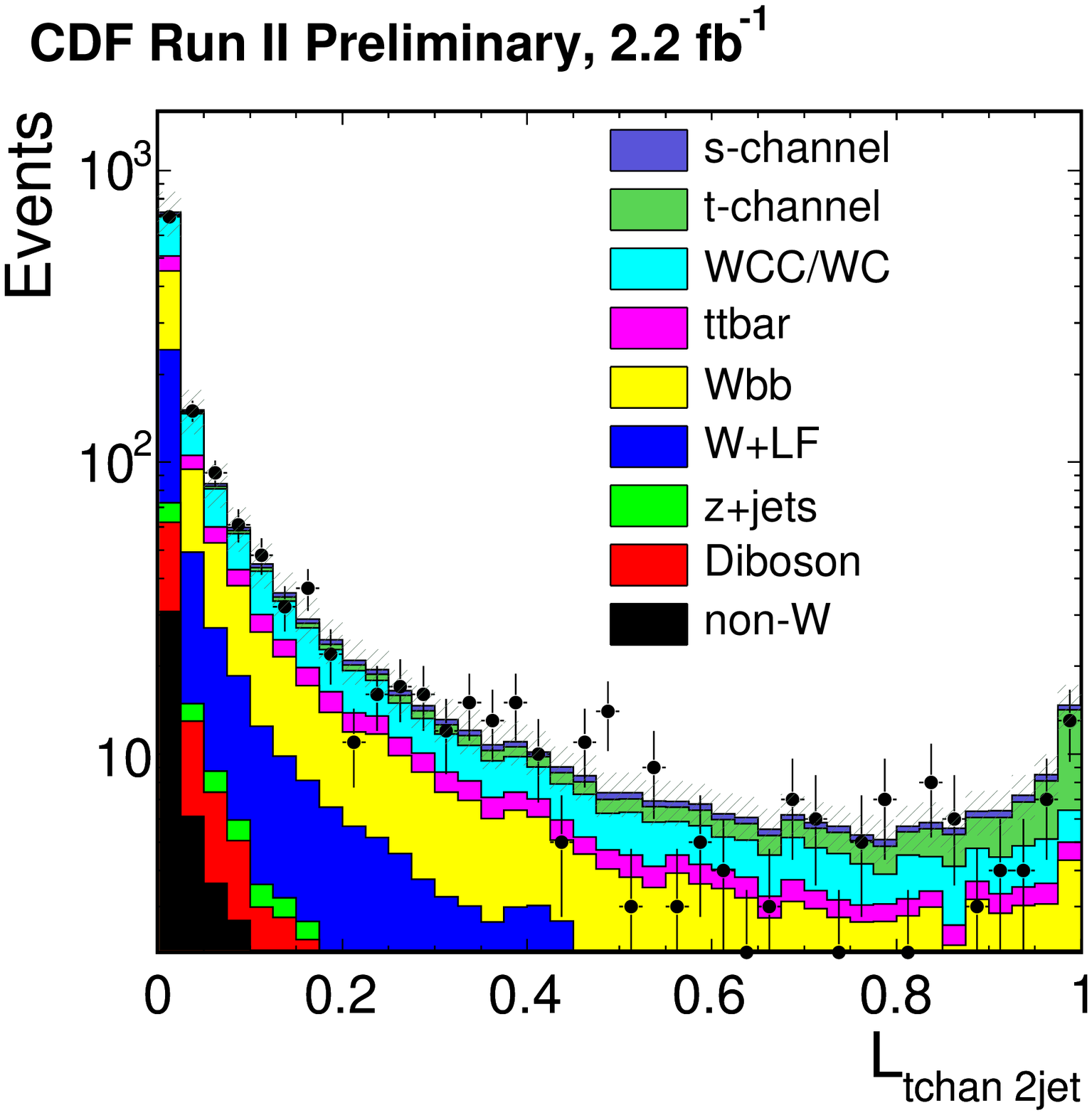,width=2.05in}
  \psfig{figure=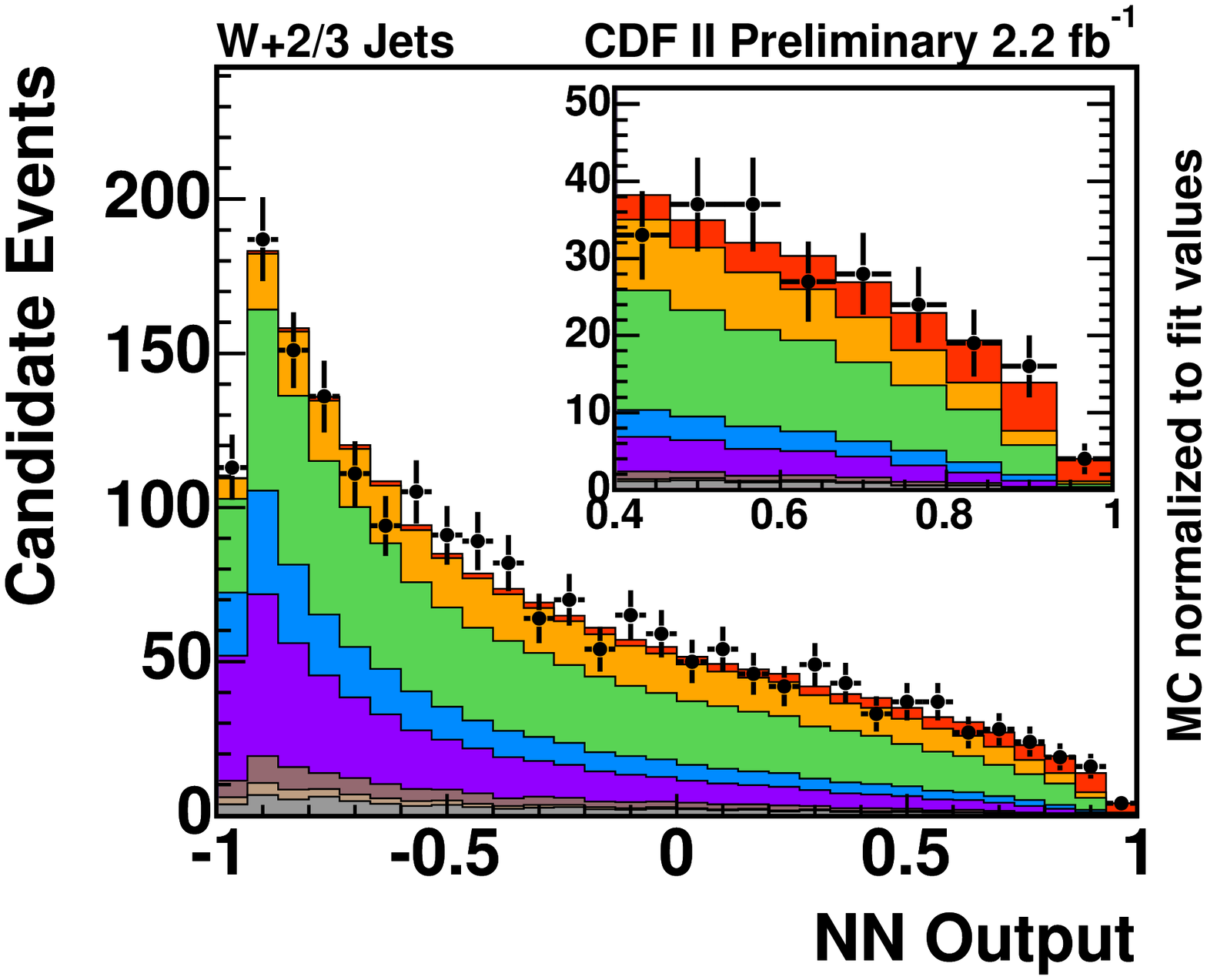,width=2.05in}
  \psfig{figure=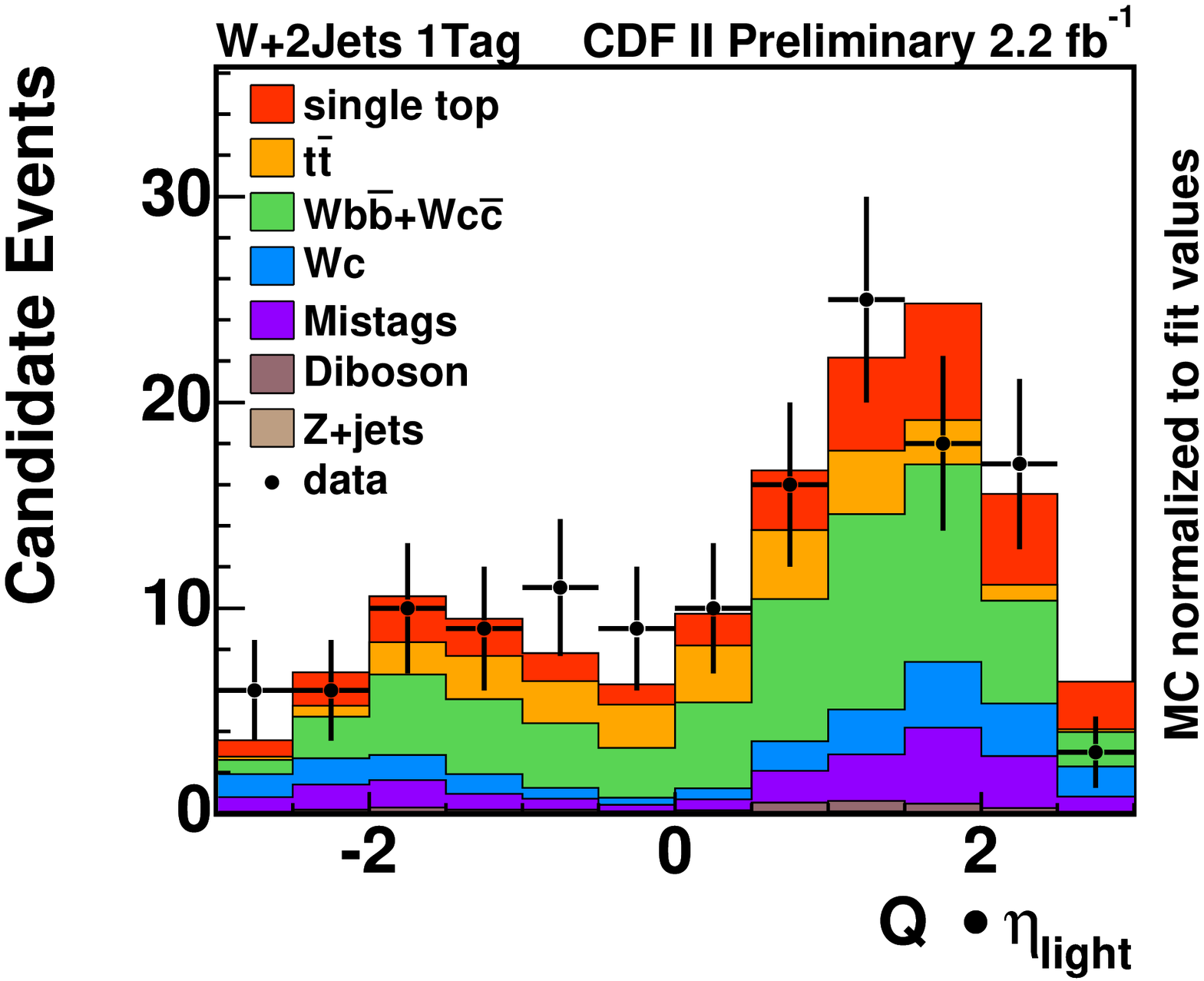,width=2.05in}
\caption{Comparison between data and background sum for the t-channel 
likelihood discriminant (left), the neural network discriminant (center),
and the light quark jet pseudorapidity in the high-discriminant region for the
neural network analysis (right). 
The signal has been normalized to the SM expectation. 
\label{fig:cdf-lhoutput}}
\end{figure}

A multivariate likelihood is built from several kinematic variables that each separate the
single top quark signal from the backgrounds.
One special variable is a specially developed b-tagging neural network that aids in separating 
b-quark jets from light quark and c-quark jets. An additional special variable is a kinematic
solver using constraints from the $W$~boson mass and the top quark mass to determine if an
event is well reconstructed. Another special variable is the t-channel matrix element, which
uses the kinematic information provided by the kinematic solver. The likelihood discriminant
for the t-channel likelihood is shown in Fig.~\ref{fig:cdf-lhoutput} (left). 

The measured cross
section is obtained as the peak of a Bayesian posterior probability. The likelihood analysis
measures a cross section of $\sigma(tb+tqb)=1.8^{+0.9}_{-0.8}$~pb, below the SM expectation.


\subsection{CDF Neural Network}
Several kinematic variables as well as the b-tagging neural network output are combined in 
a neural network. Four different networks are built with 10-14 variables each, 
trained separately for 2-jet and 3-jet as well as 1-tag and 2-tag events. The full neural
network output distribution is shown in Fig.~\ref{fig:cdf-lhoutput} (center), and the
signal region is shown in Fig.~\ref{fig:cdf-lhoutput} (right).
The neural network analysis measures a cross section of 
$\sigma(tb+tqb)=2.0^{+0.9}_{-0.8}$~pb, below the SM expectation but consistent with the
SM within uncertainties.

\subsection{CDF Matrix Element}
The matrix element analysis uses the same approach as described above, but also includes
a top pair matrix element in the 2-jet bin. The matrix element for top quark pair events 
has more final state
particles than the single top process, and these additional particles have to be integrated
out. This is done by integrating over the kinematics of the hadronically decaying $W$-boson
in a lepton+jets top pair event.  

\begin{figure}
  \centering
  \psfig{figure=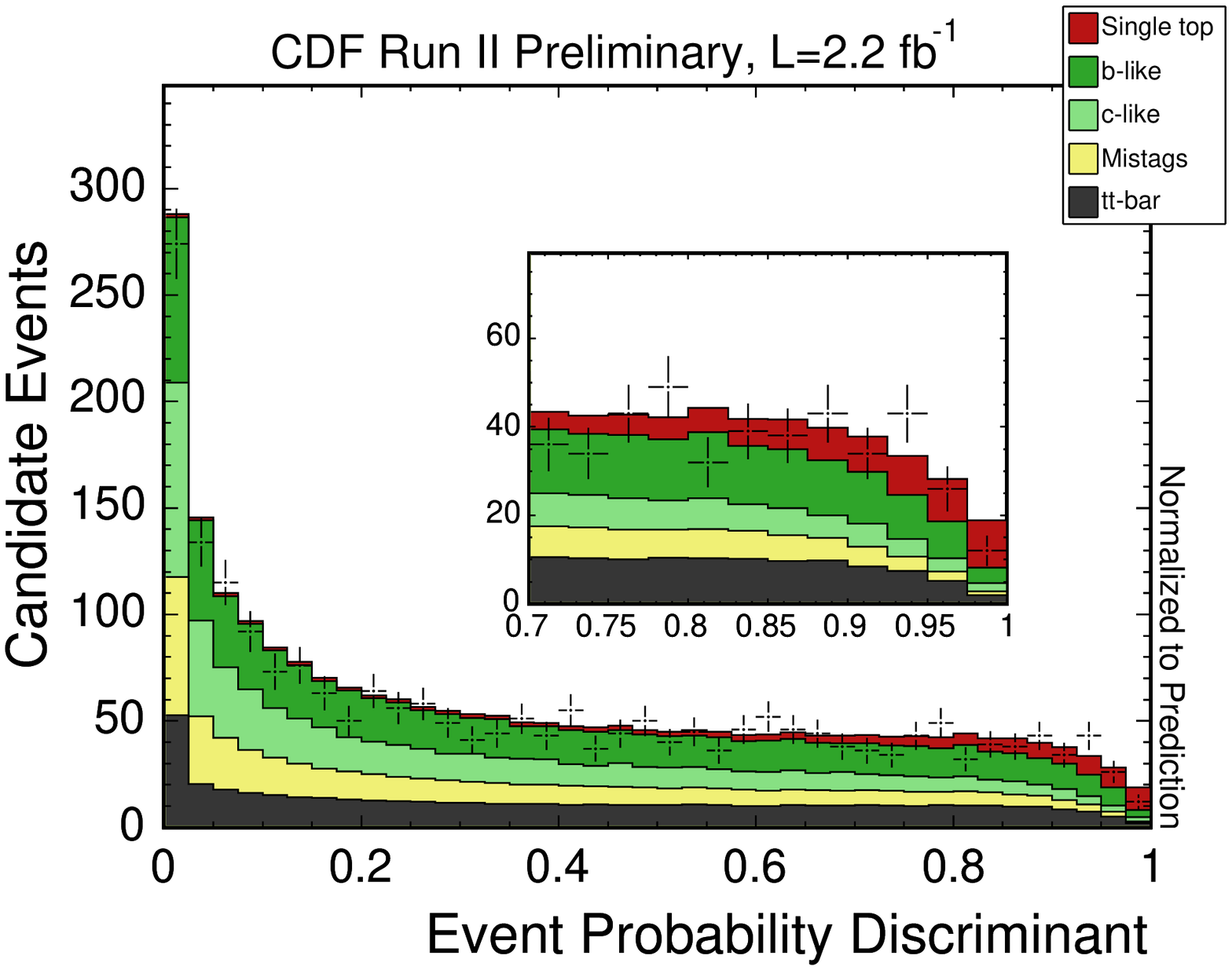,height=2.2in}
  \psfig{figure=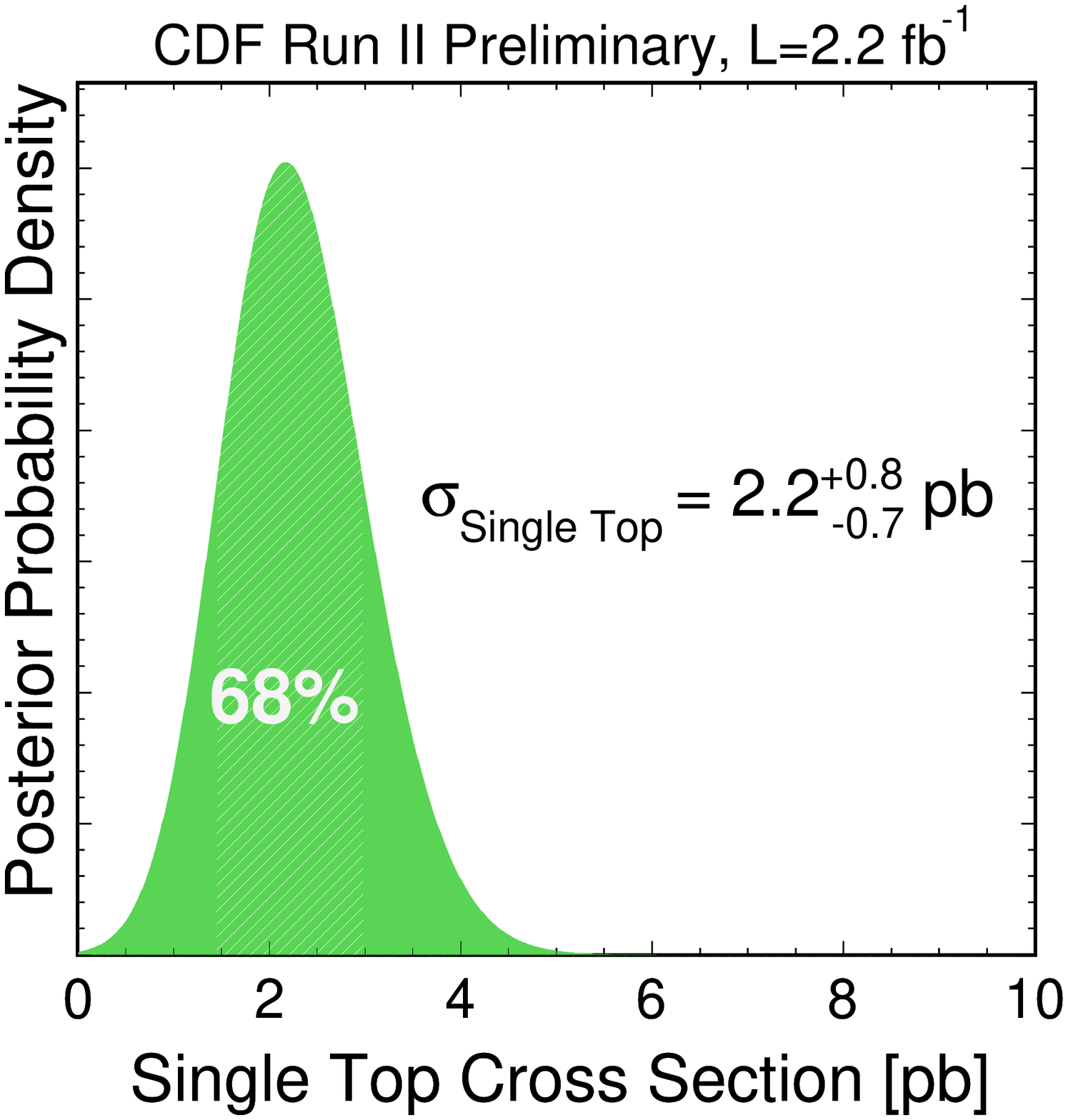,height=2.2in}
\caption{Data-background comparison for the matrix element discriminant (left) and
Bayesian posterior density distribution observed by the Matrix element analysis. 
\label{fig:cdf-me-posterior}}
\end{figure}
The Bayesian posterior probability density for the Matrix element analysis is shown
in Fig.~\ref{fig:cdf-me-posterior}, showing the measured cross section and the measurement 
uncertainty. The mesured cross section is 
$\sigma(tb+tqb)=2.2^{+0.8}_{-0.7}$~pb, again below the SM expectation but consistent with the
SM within uncertainties.
The CKM matrix element $\vtb$ is also extracted from the posterior probability and a lower limit
is found to be $\vtb>0.59$ at the 95\% confidence level.


\section{Summary}
Both Tevatron experiments have found better than 3 sigma evidence for single top quark production
and have made the first direct measurement of the CKM matrix element $\vtb$ using
advanced multivariate techniques. The CKM matrix element $\vtb$ can be measured to better than
15\%.
Further improvements to the analyses are in progress and both experiments are working towards
observation of single top quark production at the 5 sigma level.

\section*{Acknowledgments}
We thank the Fermilab staff and the technical staffs of the participating institutions for their
vital contributions.


\end{document}